# Mapping the nanoscale optical topological textures with a fiber-integrated plasmonic probe

Yunkun Wu[#1,2,3], Shu Wang[#1,2,3], Xinrui Lei[#4], Jiahui Mao[1,2,3], Liu Lu[5], Yue Liu[1,2,3], Guangyuan Qu[6], Guangcan Guo[1,2,3], Qiwen Zhan[4, †], Xifeng Ren[1,2,3, ‡]

[1]*CAS Key Laboratory of Quantum Information, University of Science and Technology of China, Hefei 230026, China*

[2]*CAS Synergetic Innovation Center of Quantum Information & Quantum Physics, University of Science and Technology of China, Hefei 230026, China*

[3]*Hefei National Laboratory, University of Science and Technology of China, Hefei 230026, China*

[4]*School of Optical-Electrical and Computer Engineering, University of Shanghai for Science and Technology, Shanghai 200093, China*

[5]*School of Mechanical Engineering, Jiangsu University, Zhenjiang 212013, China*

[6]*School of Physical Sciences, University of Science and Technology of China, Hefei, Anhui 230026, China*

[#]These authors contributed equally to this work.

[†]qwzhan@usst.edu.cn

[‡]renxf@ustc.edu.cn

**Abstract**: Topologically protected quasiparticles in optics have received increasing research attention recently, as they provide novel degree of freedom to manipulate light-matter interactions and exhibiting excellent potential in nanometrology and ultrafast vector imaging. However, the characterization of the full three-dimensional vectorial structures of the topological texures at the nanoscale has remained a challenge. Here, we propose a novel probe based on the fiber taper-silver nanowire waveguide structure to achieve super-resolution mapping of the topological textures. Based on the mode selection rules, the three-dimensional decomposed electric fields in both the far-field and near-field are directly collected and reconstructed without postprocessing algorithms, clearly visualizing the topological texures formed in free space and evanescent waves respectively. The fiber-integrated probe is further demonstrated to be robust and broadband. This approach holds promise for the characterization of more sophisticated topology in optical field, which may allow for advance applications in optical information processing and data storage.

Topologically protected quasiparticles characterized by nontrivial spin textures emerge in many areas ranging from high-energy to condensed matter physics [1-3]. As a prominent example, skyrmions in magnetic materials have attracted numerous research attention for the past decades due to the ultracompact size and great stability [3-5], promising for data storage and information processin [6,7]. The topological feature was observed in optical system recently [8-9], with the skyrmionic textures constructed by either electromagnetic fields [8,10-11], spin angular momentums [9,12-15] or Stokes vectors [16-18]. The topological stability and deep-subwavelength properties of optical skyrmions provide new degrees of freedom in optical manipulation and light-matter interaction at nanoscale, enabling many applications in nanoscale metrology [19,20], ultrafast vector imaging [10,21] and topological Hall devices [22].

While numerous topological textures have been proposed in optics, the characterization of sophisticated vectorial structures is still challenging, especially in the presence of high inhomogeneous of light in focusing beams and evanescent waves. Near-field scanning optical microscope (NSOM) with nanostructure-based probes has been employed to visualize optical skyrmions experimentally, where longitudinal electric field can be measured by scattering-type NSOM with pseudo-heterodyne interferometric detection [8], and transverse optical fields are obtained by aperture-based probes [9,23]. Photoemission electron microscopes are also utilized to provide spatiotemporal vector dynamics of optical skyrmions [10,21]. However, only the electromagnetic field component parallel or normal to the interface was measured in the previous approach due to the single resonance of probe, and other orthogonal polarized components are obtained from Maxwell's equations.

The access to the complete three-dimensional(3D) decomposed information of vectorial optical field with subwavelength resolution is of crucial significance for the characterization of photonic quasiparticles, but remains a challenge. In general, the realization of full vectorial imaging of optical field relies on the detection of far-field scattered light from metal/dielectric particles or tips [24-28], where the probe works solely as a scatterer rather than a collector, necessitating an additional non-integrated collection optical path. And the scattering light is frequently sensitive to the morphological parameters of the probe, which poses difficulty in characterizing multiple topological textures formed in different scenes. Single photon emitters (SPEs), such as nitrogen-vacancy color center in diamonds, have also been used as a probe to measure all components of vector optical field based on the interaction between the light field and the emitters [29-31]. But each SPE could only effectively characterize a specific optical polarization direction contingent on the orientation of SPE itself, requiring multiple repeated measurements through SPEs with different well-defined axes.

In this work, we propose a plasmonic probe based on the fiber taper-silver nanowire (AgNW) hybrid waveguide, which possesses the capability to characterize the photonic topological textures formed in both near- and far-field. By analyzing the hybrid eigenmode inside the AgNW, we demonstrate the polarization-selective coupling of plasmonic modes, which can be utilized to measure each orthogonal electric field component. The nanoscale 3D-decomposed vector optical field is mapped directly by collecting the transmitted light through the fiber-integrated free-standing probe in experiments, offering high flexibility for applications in the confined spaces without the need for reconstruction algorithms. Topological textures formed in free space and evanescent waves are visualized respectively. The proposed probe is also demonstrated to have good robustness of preparation and a multiwavelength working bandwidth, enabling further integration with spectral techniques to detect more kinds of vector optical signals, including photoluminescence, linear and nonlinear optics, etc.

The decomposition and detection of the vectorial light field using the hybrid probe rely on the selective excitation and coupling of specific modes. A schematic diagram is shown in Figure 1. The x and y directions are defined as perpendicular to the AgNW, while the z direction is parallel to the AgNW. The dominant optical mode supported by an elongated fiber is a hybrid eigenmode (HE), encompassing two degenerate orthogonal modes denoted as $LP_{1a}$ and $LP_{1b}$ [32]. For a suspended AgNW, the two lowest-order surface plasmon polariton (SPP) eigenmodes are the fundamental transverse magnetic (TM) mode, denoted as $H_0$, and the second-order hybrid mode (HE) comprising two degenerate modes, denoted as $H_{1a}$ and $H_{1b}$ respectively [33].

The whole process of light collection by the probe can be delineated into three steps: excitation, coupling and measurement. Initially, each electric field component $E_x$, $E_y$, $E_z$ at the center of the AgNW end would respectively excite $H_{1a}$, $H_{1b}$ and $H_0$ SPP mode on the AgNW, owing to the mode field matching (see Fig. S1 in the Supplement). Subsequently, these SPP modes would selectively couple to specific optical mode on the fiber taper based on the coupled mode theory, which has been demonstrated in our prior works [34,35]. As the fiber taper and AgNW are aligned along the y-axis (Figure 1a), $H_{1a}$ mode exclusively couples to $LP_{1a}$ mode, while both $H_{1b}$ and $H_0$ modes are anticipated to couple to $LP_{1b}$ mode. A 90-degree rotation of the probe would maintain the coupling behavior for HE modes, while the fundamental mode $H_0$ couples to another fiber mode $LP_{1a}$ (Figure 1b). Finally, the $LP_{1a}$ and $LP_{1b}$ modes in fiber are segregated by polarizing optical elements and measured with each intensity containing the information of the collected electric field, which enables 3D-decomposation of vectorial optical field.

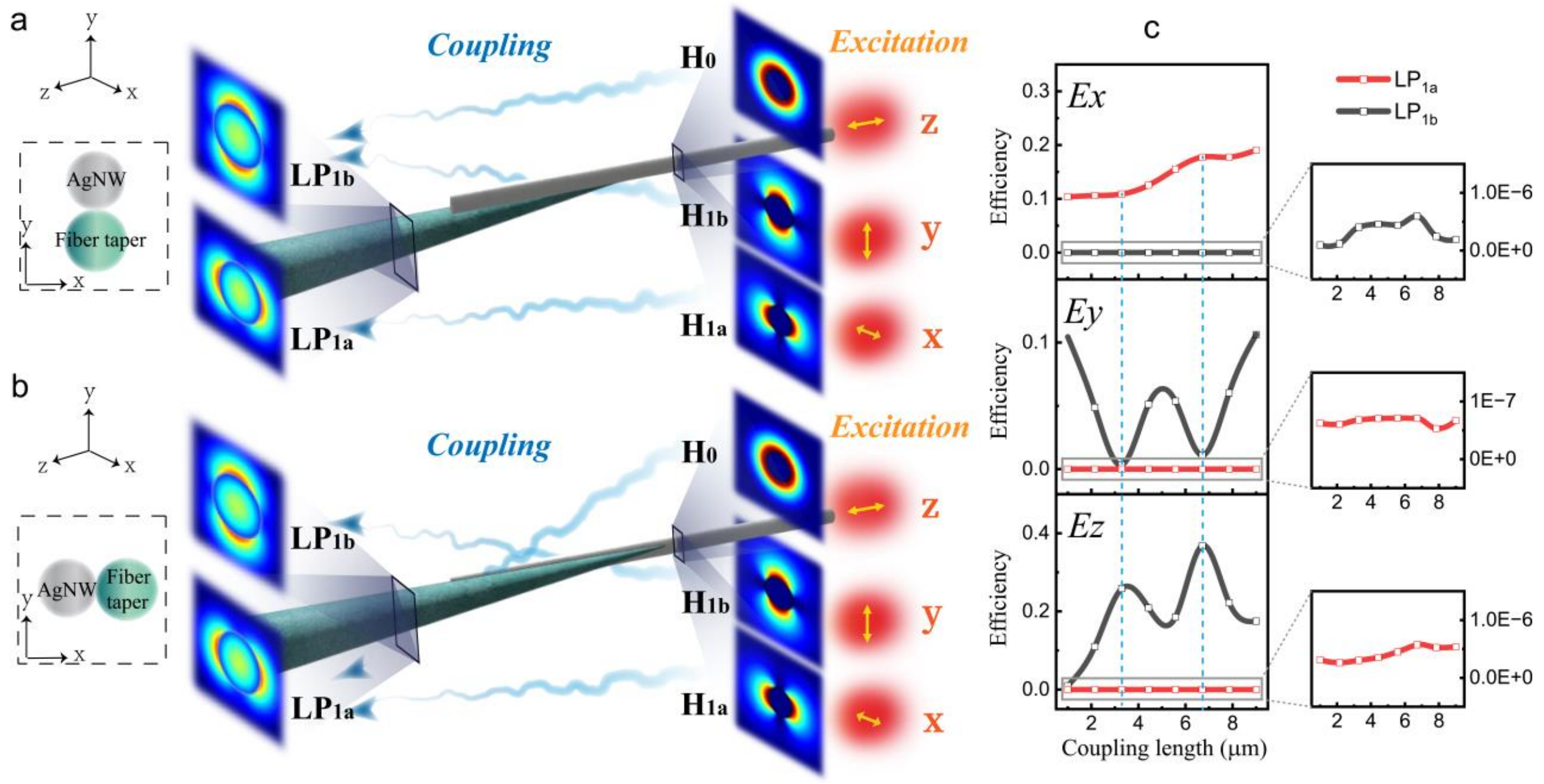

**Fig. 1: Principle of the hybrid probe for measuring decomposed vector light fields. a** $H_0$ and $H_{1b}$ SPP modes couple to the $LP_{1b}$ optical mode in the fiber taper, while $H_{1a}$ SPP mode couples to the $LP_{1a}$ optical mode in the fiber taper, when fiber taper and AgNW are aligned along the y-axis. **b** $H_{1b}$ SPP mode couples to the $LP_{1b}$ optical mode in the fiber taper, while $H_0$ and $H_{1a}$ SPP modes couple to the $LP_{1a}$ optical mode in the fiber taper, when fiber taper and AgNW are aligned along the x-axis. **c** The simulated efficiency of the probe detecting $E_x$, $E_y$, $E_z$ versus coupling length for the situation in **a**. Red and black lines represent the efficiency of converting to $LP_{1a}$ and $LP_{1b}$ mode respectively, which are interchanged when the situation is **b**. Blue dashed lines point out the chosen coupling length of the probe in experiments.

For the case in Figure 1a, two sets of intensity results can be obtained as $a|E_x|^2$ and $\left|\sqrt{b}E_y + \sqrt{c}E_z\right|^2$ from the fiber taper, where coefficients $a$, $b$, $c$ represent the detection efficiency of $E_x$, $E_y$, $E_z$, respectively. The structural parameters of the hybrid waveguide do not affect the aforementioned mode selection rules of excitation and coupling, while effectively modulating the efficiencies [36]. Particularly, the detection efficiency of $H_0$ to $LP_{1b}$ exhibits an inverse trend

relative to that of $H_{1b}$ to $LP_{1b}$, as shown in Figure 1c. Hence each electric field component can be extracted by optimizing the probe parameters. By adjusting the coupling length to the value marked by the blue dashed lines in Figure 1c, the coefficients realize $b/c \leq 0.01$ and the approximation $\left|\sqrt{b}E_y + \sqrt{c}E_z\right|^2 \approx c|E_z|^2$ can be established. Similarly, after a 90-degree rotation of the probe as depicted in Figure 1b, intensities of $a\left|E_y\right|^2$and $\left|\sqrt{b}E_x + \sqrt{c}E_z\right|^2 \approx c|E_z|^2$ can be obtained. Therefore, the full vector components of the light field can be detected after being scanned twice by the probe without additional algorithms, and the mapping of $E_x$ and $E_y$ is automatically normalized thanks to the structure symmetry. The distributions of $E_z$ are normalized by calibration with theoretical strength in advance, and are presented as the average of the two results before and after rotation.

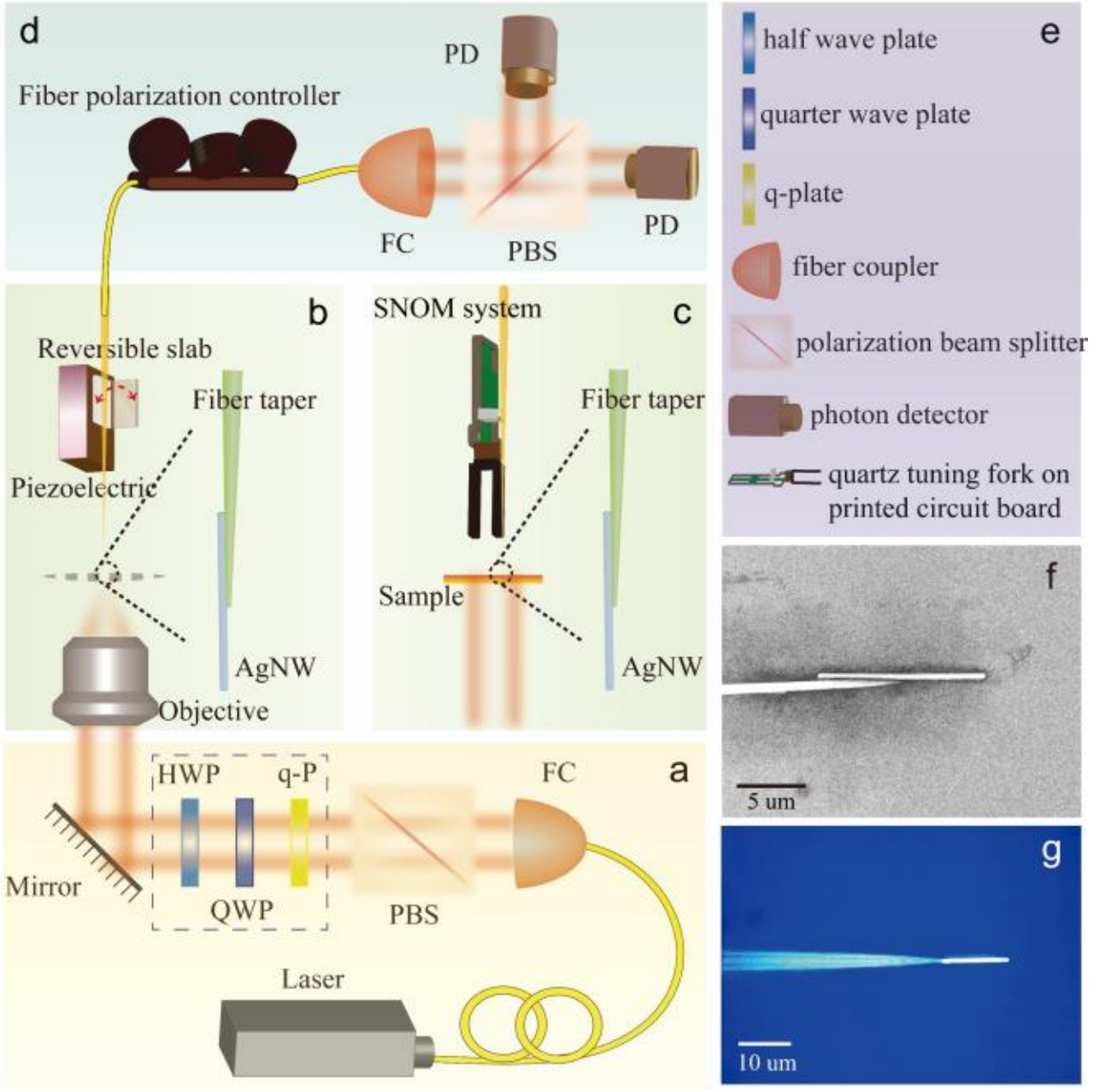

**Fig. 2: Experimental setup.** **a** Light source of the experiments. Incidence with different polarizations is prepared through selectively incorporating wave plates in the dashed box. **b** Far-field experiment setup, where the probe measures tightly focused light beams with different polarizations. **c** Near-field experiment setup, where the probe measures optical field distributions in the near-field of the samples under illumination. **d** Decomposition and detection part of the experiments. $LP_{1a}$ and $LP_{1b}$ modes in the fiber are separated by PBS and measured by photon detectors. **e** Introduction of abbreviations in the schematic diagram. **f** SEM of the probe. **g** CCD image of the probe.

To demonstrate the efficacy for characterizing the photonic topological textures with the proposed hybrid probe, the 3D-decomposed vector light fields in both far and near field are experimentally investigated. The corresponding diagram of the experiment setup is shown in Figure 2. Far-field experiments were tailored to measure the optical distributions of a tightly focused optical field with varied polarizations, whereas near-field experiments aimed to measure the near-field optical distributions with the interaction between laser beam and metallic nanostructures.

Figure 2a illustrates the light source of the experiments for both far-field and near-field, where the laser could be modulated into linear, circular, azimuthal, and radially polarizations with the waveplates in the dashed box. The sensing parts for far-field and near-field experiment are shown in Figures 2(b-c) respectively. In the far-field experiments, the collimated laser with different polarizations was focused by an objective (NA=0.8), and the focal plane was scanned by the probe, which was placed on a reversible slab rotatable by 90°. While in the near-field, the collimated laser modulated to the desired polarization was directly illuminated onto the sample without focusing. The probe affixed to a quartz tuning fork was assembled into the NSOM system to measure the near-field above the surface of samples. The decomposition and detection part for the far-field and near-field experiments were the same as depicted in Figure 2d. A fiber polarization controller (FPC) maintained the polarization state of the photons after passing through the fiber, and a second PBS divided $LP_{01}$ and $LP_{02}$ modes. The scanning electron microscope (SEM) and optical image of our probe are exhibited in Figures 2(f-g).

The 3D-decomposed distributions of the tight focused optical beam with different polarizations are shown in Figure 3. An 808nm wavelength laser with three kinds of polarizations was prepared and focused for measurement. The structural parameters of our probe in experiment are optimized as AgNW radius of 280 nm, coupling length of 6.2 $\mu m$ and cone angle of 3.6 degree from the simulation results. For x-linear polarized illumination, the focused field only contains x and z components, as shown in Figure 3a. Due to the different interfering behavior near the focus for each component, a bright spot is demonstrated at the center of $E_x$, while the intensity of longitudinal field $E_z$ reaches minimum at the vicinity of y axis and opposite sides of the beam are out of phase. This gives rise to the formation of one dimensional skyrmions along x axis, where the local electric vector points along negative z axis at the boundary with $E_x$=0, and rotates progressively along positive x axis with a reversal to ‘up’ state at another boundary.

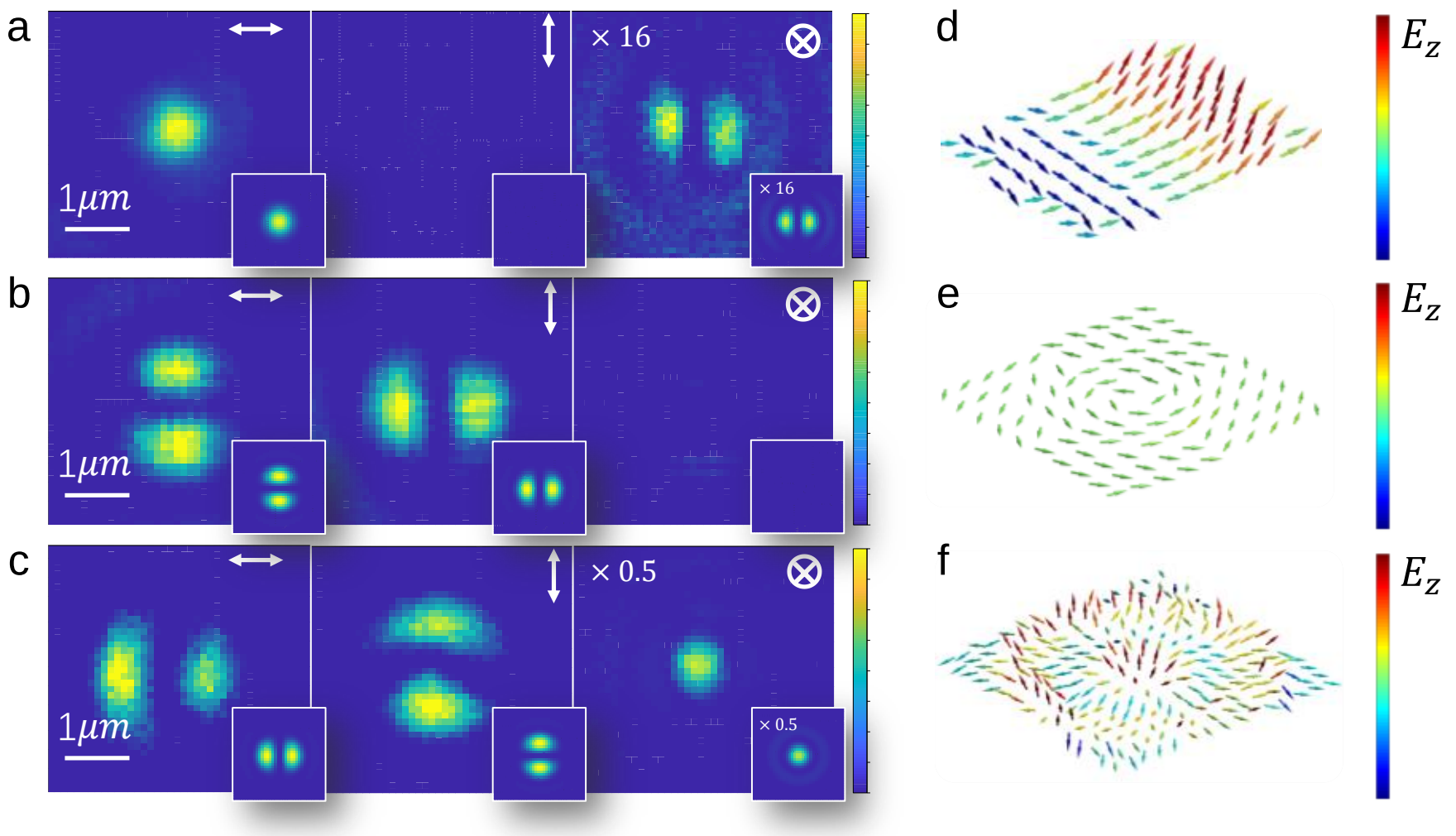

**Fig. 3: Results of 3D-decomposed optical field distributions and topological textures of the tight focused optical beam with different polarizations. a-c** The experimental reconstructed $E_x$, $E_y$, $E_z$ of the tight focused optical beam with inset denoting the theoretical results for **a** x-linear polarization, **b** azimuthal polarization and **c** radially polarization at 808nm wavelength, respectively. White arrows in the upper right corner of each image indicate the polarization-decomposed components of the vectorial optical field, corresponding to $E_x$, $E_y$, $E_z$ from left to right. **d-f** Normalized electric vectors depicted according to **a-c**, with the arrows indicating the direction of electric vector.

For azimuthally polarized incidence, only the azimuthal component is present at the focus with donut shape, and the corresponding x and y components consist of two lobes (Figure 3b). The local electric vectors form a topological vortex rotating along xy plane. While for radially polarized incidence, only the radial and longitudinal components are present at the focus, with donut and bright spot obtained respectively (Figure 3c). In contrast, the longitudinal components are lost for the traditional fiber-based probes as shown in Fig. S5, which fails to characterize the focusing property. The intensity around the center lobe of $E_z$ decrease rapidly due to the high numerical aperture, and the phase is opposite on the two sides of minimum intensity. Hence, the local electric vector varies progressively from the central "up" state along the radial direction, resulting in a skyrmion-like structure (Figure 3f). The corresponding theoretical results of focal field in each polarization calculated by Richards-Wolf vectorial diffraction method are shown in the inset of Figure 3, which agree well with experimental measurement.

To demonstrate the structural robustness of our probe which is crucial for stable properties and reproducible production, further simulations involving a detailed sweep of the coupling length around the selected value, the radius of the AgNW, the cone angle of the fiber taper and the wavelength are shown in Fig. S1 and Fig. S2. Although our proposal places certain constraints on the coupling length, it allows for an 800 nanometers error margin, which is controllable enough within high magnification optical microcopy practices. Additionally, AgNWs with diameters ranging from 250nm to 400nm and fiber taper with cone angles from 2.7 to 5.2 degrees yield satisfactory results. The simulation result without dramatic change shows that neither the excitation nor the coupling process is resonant in our proposal, indicating not only the aforementioned robustness but also a multiwavelength operating bandwidth. This distinguishes our probe from some previous works relying on SPP resonant modes [31]. The focused radially polarized laser with wavelengths of 633 nm and 1020nm are also measured in Fig. S3, showcasing the broadband capability of the probe. More details and discussions on the robustness and work bandwidth of the probe can also be found in the Supplementary information. Furthermore, to address the oxidation of the AgNW in air, we coated the AgNW with a layer of silica that was approximately several nanometers thick using the chemical method. In this way, after placing the probe in air for a month, the coated probe still showed a good efficiency and sensitivity, as demonstrated in Fig. S7.

The sensing of the near-field topological textures with our probe was also experimentally studied. After attached to the quartz tuning fork, the probe can accurately reach the surface of the samples and stay steady above it with the mature shear force feedback control mode. Therefore, the hybrid probe can function effectively as a practical NSOM probe, offering some advantages over commercial counterparts. In addition to the ability to collect the in-plane and out-of-plane components simultaneously, the experimental optical transmission efficiency, including the focusing efficiency of the fiber taper, the coupling efficiency of the hybrid region and the propagation loss on the AgNW, exceed 29% for each linearly polarized incidence (Fig. S8). This provides high signal to noise ratio and robustness for the measurement. In contrast, the typical transmission efficiency of the commercial NSOM probes is approximate 0.01%. An even higher transmission efficiency of 70% can be achieved by modulating the parameters of this hybrid waveguide according to reference [37].

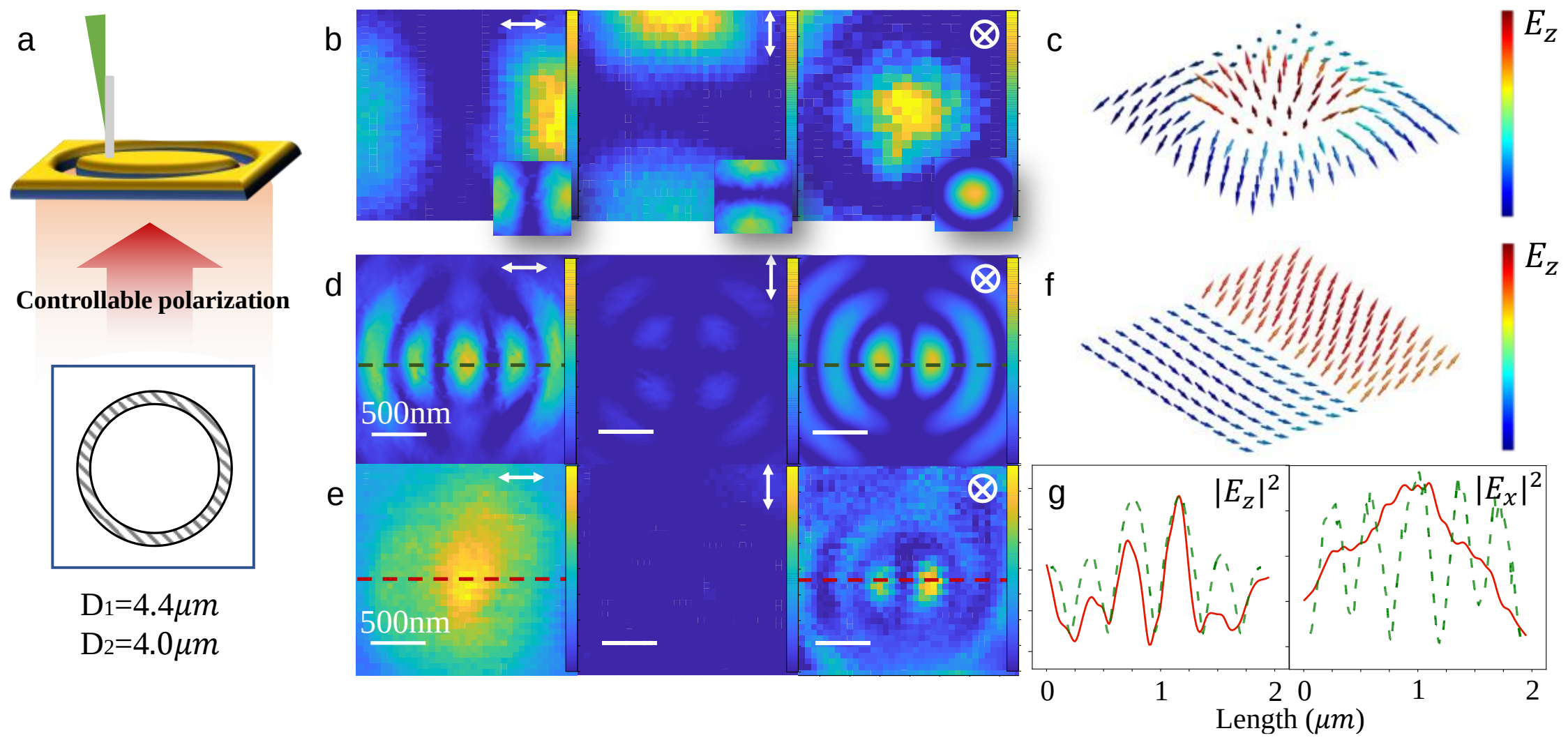

**Fig. 4: Characterization of near-field 3D-decomposed optical field distributions with metal nanostructures under illumination. a** A nanoring groove etched on the Au film was illuminated by laser beam, and the excited SPP field on the surface of the Au film was measured by the probe. **b** Experimental measurement of each plasmonic field component under radially polarized illumination, with simulation results shown in the insets. **c** Neel type electric field skyrmion formed in b, with the arrows indicating the direction of electric vector. **d** Theoretical and **e** experimental results of plasmonic fields under linear polarized illumination. **f** Normalized electric vectors in e, forming one dimensional skyrmions along x axis. **g** Crosscuts of the intensities for $E_z$ and $E_x$ along the dashed lines in **d** and **e**. Red solid lines and green dashed lines in represent theoretical and experimental data, respectively. White arrows in the upper right corner in b, d and e indicate the polarization-decomposed field components $E_x$, $E_y$, $E_z$ from left to right.

To generate near-field topological textures, laser beam with a wavelength of 808 nm is illuminated onto a metallic structure, which is comprised of a nanoring groove etched into a 150 nm gold film with a glass substrate (Figure 4a). The incidence radiated on the groove would excite SPP on the sidewall of the groove [38], thus each point on the sidewall of the ring groove could be regarded as a secondary source of SPP to generate interference fringes inside the ring. For radially polarized incidence, the plasmonic field excited on each sidewall is homogenous and would converge to the center of the nanoring. Evanescent Bessel beams with first and zero orders are generated in radial and longitudinal field components (Figure 4b), which is different with the focused radially polarized beam in free space where the electric field near the focus is an integral of Bessel beams with different in-plane wave vectors. Numerical simulation agrees well with experimental results (insets in Figure 4b). The evanescent feature of plasmonic field gives rise to a smaller focal spot in $E_z$, and the electric vectors vary progressively from the central “up” state to the edge “down” state, manifesting a Neel-type photonic skyrmion (Figure 4c and Fig. S10).

While for x-linear polarized illumination, only the radial component can couple to and excite SPPs, leading to different excitation efficiency on each sidewall. Constructive and destructive interference would occur at the center of the ring for in-plane and out-of-plane SPP respectively, as simulated in Figure 4d. The corresponding experimental results measured by the probe are shown in Figure 4e. Similar to the focused linear polarized beam in free space, the central bright and dark spots in $E_x$ and $E_z$ in focused plasmonic field form one dimensional skyrmions along x axis with varying electric vector, while the size is smaller due to the evanescent feature of SPP (Figure 4f). Crosscuts of the intensity of $E_x$ and $E_z$ along the dashed lines in Figures 4(d-e) are depicted in Figure 4g, wherein the experimental distributions are drawn in red and the theorical distributions are drawn in green. A resolution of 224nm was achieved for $E_z$, which breaks the diffraction limit of light. However, the in-plane electric field $E_x$ failed to reach the same resolution, indicating higher resolution for our probe in the out-of-plane optical field compared to the in-plane optical field. This anisotropy can be attributed to the different effective mode areas of $H_0$ and $H_1$ modes, as calculated in Figure S9. In general, higher resolution (or smaller effective mode area) requires a smaller AgNW radius. However, the second-order $H_1$ mode becomes divergent with a radius below the threshold, while the fundamental $H_0$ mode does not have a cutoff radius. Therefore, achieving even higher resolution for the out-of-plane optical field is feasible with a thinner AgNW.

In addition, the topological features for localized surface plasmons (LSPs) generated on a metallic hollow circular hole is also measured with the probe. Collimated circularly and linearly polarized laser were illuminated on the nanostructure, and the measured pronounced longitudinal component along the edge of the hole clearly revealed the excitation of LSPs (Figs. S11-12). Spatial resolution of 373nm and 358nm were achieved simultaneously for in-plane field ($E_x$, $E_y$) and out-of-plane field ($E_z$) respectively, breaking the diffraction limit for each polarization component.

In summary, we have proposed and demonstrated an efficient method for mapping the photonic topological texture through a hybrid probe, where the 3D-decomposed vector optical field is measured with a resolution surpassing the optical diffraction limit. Based on the excitation and coupling mode selection rules, the waveguide-based probe enables direct decomposition and collection of the orthogonal components of electric field, reconstructing each component without postprocessing algorithms. Experimental measurements of the topological textures formed in tightly focused light field in the far-field and SPP at near-field have been investigated, both of which agree well with theoretical predictions. This straightforward and efficient approach, made possible by the highly effective, fiber-integrated, stable, and broadband nature of the probe, holds significant potential for studying nanoscale light-matter interactions and complex artificial light fields. On the other hand, this method can also be used to excite controllable vector optical field at naoscale. The probe served as a nanoscale collector and light source for 3D vector optical field, offers a valuable tool to advance applications of 2D materials, biosensing, superresolution imaging, optical communication, etc.

## Acknowledgements

This work was supported by National Key Research and Development Program of China (2022YFA1204704), the Innovation Program for Quantum Science and Technology (grant number 2021ZD0303200), the National Natural Science Foundation of China (NSFC) (62061160487, T2325022, U23A2074, 62205325, 12104440, 92050202, 12204309), the CAS Project for Young Scientists in Basic Research (No. YSBR-049), Key Research and Development Program of Anhui Province (2022b1302007) and the Fundamental Research Funds for the Central Universities. This work was partially carried out at the USTC Center for Micro and Nanoscale Research and Fabrication.